\documentclass[preprint,superscriptaddress]{revtex4-1}
\pdfoutput=1
\usepackage{amsmath, amssymb}
\usepackage{graphicx}
\usepackage{color}
\usepackage{hyperref}
\newcommand{\textmark}[1]{\textcolor{black}{ #1}}

\begin{document}

\title{\bfseries Information density, structure and entropy in equilibrium and non-equilibrium systems}

%{Comparing different approaches to estimate non-equilibrium entropy}

\author{Mengjie Zu}
\email[(Equal Contribution) ]{mz410@cam.ac.uk}
\affiliation{Institute of Physics, Chinese Academy of Sciences, 8 Third South Street, Zhongguancun,
Beijing 100190, China.}
\affiliation{Department of Chemistry, University of Cambridge, Cambridge, UK.}
\author{Arunkumar Bupathy}
\email[(Equal Contribution) ]{bupathy@jncasr.ac.in}
\affiliation{Theoretical Sciences Unit, Jawaharlal Nehru Centre for Advanced Scientific Research, Bangalore, India.}

\author{Daan Frenkel}
\email{df246@cam.ac.uk}
\affiliation{Department of Chemistry, University of Cambridge, Cambridge, UK.}

\author{Srikanth Sastry}
\email{sastry@jncasr.ac.in}
\affiliation{Theoretical Sciences Unit, Jawaharlal Nehru Centre for Advanced Scientific Research, Bangalore, India.}
%\footnote{MZ and AB contributed equally to this work.}
\begin{abstract}
{\footnotesize During a spontaneous change, a macroscopic physical system will evolve towards a macro-state with more realizations. This observation is at the basis of the Statistical Mechanical version of the Second Law of Thermodynamics, and it provides an interpretation of entropy in terms of probabilities. However, we cannot rely on the statistical-mechanical expressions for entropy in systems that are far from equilibrium. In this paper, we compare various extensions of the definition of entropy, which have been proposed for non-equilibrium systems. It has recently been proposed that measures of information density may serve to quantify entropy in both equilibrium and nonequilibrium systems. 
We propose a new ``bit-wise'' method to measure the information density for off-lattice systems. This method does not rely on coarse-graining of the particle coordinates. We then compare different estimates of the system entropy, based on information density and on the structural properties of the system, and  check if the various entropies are mutually consistent and, importantly, whether they can detect non-trivial ordering phenomena. 
We find that, except for simple (one-dimensional) cases, the different methods yield answers that are at best qualitatively similar, and often not even that, although in several cases, different entropy estimates do detect ordering phenomena qualitatively. % \DFcomment{I do not think that this addition greatly improves the text.}\\
Our entropy estimates based on bit-wise data compression contain no adjustable scaling factor, and show large quantitative differences with the thermodynamic entropy obtained from equilibrium simulations.  Hence, our results suggest that, at present, there is not yet a single, structure-based entropy definition that has general validity for equilibrium and non-equilibrium systems.}
    
%AB: Information density works pretty well when there is enough difference in the structural properties of the different states being compared. It falls flat on its face when the differences in the states are more subtle. S2, on the other hand, does a better job of detecting the transitions.

%MZ: S_2 does better works for liquid states, but we should notice that it does not work for solid states, so I doubt that S_2 could detect transitions better than other methods. %AB: I don't see the supporting data.

\end{abstract}

\maketitle
% \DFcomment{There is possibly an issue with the notation: we have $\ln$, $\log$ and $\log_2$. I do not think that $\log$ is meant to be $\log_{10}$. If not, the symbol $\log$ is confusing}
% {\color{magenta}(AB: I have updated the $\log$s to $\ln$.)}

\section{Introduction}
\label{sec:intro}
The Second Law of Thermodynamics is exceptional among physical laws in that it is usually formulated as an inequality: during a spontaneous change in a closed system, the change in entropy ($\Delta S$) cannot be negative. Since the days of Boltzmann and Zermelo, this property of entropy has given rise to endless debates. However, following Boltzmann, Gibbs and Planck, equilibrium Statistical Mechanics has established a clear link between entropy and probability, and in that framework the second law simply states that a system will spontaneously evolve towards macro-states that have more realizations than the initial macro-state, and are therefore more probable. However, for systems that are intrinsically not in equilibrium - \textit{i.e.}, where neither the initial nor the final macro-state can be described by equilibrium statistical mechanics, it is not {\em a priori} obvious how to define quantities that have one or more of the properties of entropy. 

A simple and robust definition of entropy, dating back to Gibbs, relates $S$ to the probability to find a system (equilibrium or not) in any of its micro-states, \textit{i.e.},
\begin{equation}
S = -k_B \sum_i p_i \ln p_i,
\label{eq:gibbs-entropy}
\end{equation}
where $p_i$ is the probability of the system being in a micro-state $i$, and the sum runs over all accessible micro-states.

In equilibrium, computing $S$ is  simplified by the fact that the probabilities are known \textit{a priori}. Using numerical free-energy calculation methods~\cite{Frenkel-Smit}, the computation of entropy can be extended to more complex cases. Out of equilibrium,  the probability of observing a given macro-state becomes protocol dependent and it becomes a challenge to define an entropy that increases during a spontaneous change and provides a measure for the probability of observing a macro-state. 

For certain protocols, methods such as basin-volume calculations can be used to compute a Gibbs-style entropy for systems that can be found in a large number of distinct states (e.g. stable packings of granular matter~\cite{Xu2011,Asenjo2014,Martiniani2016}). However, such calculations of the ``granular entropy'' are computationally expensive and, more importantly,  may not be not applicable to other non-equilibrium situations. It is for this reason that, in the present paper, we will not consider estimates of the entropy based on basin-volume calculations. 

A distinct way of computing the entropy of a system from knowledge of its structure, was developed in the study of fluids at equilibrium: it is based on Kirkwood's factorization of $n$-body distribution functions~\cite{Kirkwood1942}.
In this formalism, the entropy of an $n$-body system is expressed as an infinite series: $S = S_{id} + S_2 + S_3 + S_4 + \cdots$, where $S_{id}$ is the ideal gas entropy at the given density, and $S_n$ are the contribution due to $n$-body distribution functions~\cite{Green1952,Nettleton1958,Raveche1971,Wallace1987}. Although the expressions for $S_2, S_3, \cdots$ are ensemble invariant~\cite{Baranyai1989}, and can be used to compute $S$ to arbitrary precision, the numerical difficulties in computing the higher order terms reduce their practical applicability. For our purpose the interesting aspect of an entropy-estimate based on structure alone, is that it can also be used for non-equilibrium systems. However, it is not at all clear that the resulting quantity retains its entropy-like properties out of equilibrium. 

Recently, there has been a resurgence of  interest in quantifying non-equilibrium entropy in terms of  the order or the information content of a variety of physical and biological systems~\cite{Martiniani2019,Avinery2017,Melchert2015,Rams2015,Flann2013,Galas2010,Benci2004}, using the ideas of algorithmic complexity introduced by Kolmogorov and Chaitin in the context of information theory~\cite{Kolmogorov1968,Chaitin1966}. The Kolmogorov complexity $K$ is loosely defined as the length of the shortest computer program that can reproduce a given symbolic string. Unlike the  Gibbs or Shannon entropy which are statistical measures, $K$ is a measure of complexity or information content which is intrinsic to a given state, and does not require knowledge of the probability distribution of the underlying process. In the context of a many-body system, it is a quantity that can be computed for a single realization of the system. 

The interest in $K$ stems from the fact that in the limit of large string length, $K$ is proportional to the Shannon entropy~\cite{Cover2006,Shannon1948}. But given a lack of general principles to compute $K$, any real complexity analysis has to make use of a universal coding scheme, such as a lossless compression algorithm. The length of the resultant compressed string is then assumed  to be the best possible estimate of $K$, and forms the basis of \textit{computable information density} (CID) measures.

While CID measures have been used in the recent literature to identify order and information content, there is a scarcity of results that connect CID with the thermodynamic entropy. In this paper, we try to answer the question, whether there is a relation between the amount of information contained in an ensemble of micro-states as measured by the CID, and its entropy. Although entropy and order are closely related, it is important to note that statistical-mechanical entropy is actually a measure of the volume of phase space available to the system. \textmark{We also emphasize that the different methods discussed here consider only the static structure of the systems for estimating the entropy and do not consider their time evolution, even for the nonequilibrium systems studied.} Therefore, the question whether the typical ordering of particles in a given state is related to its entropy must be approached with caution, especially when out-of-equilibrium.

The key point in computing the CID of particle configurations lies in converting them to $1$-dimensional strings, which can then be compressed by the appropriate software. This is usually done in the same way as digitizing a photograph, \textit{i.e.}, the particle configurations are digitized onto a regular grid resulting in a set of occupation numbers at each grid point, corresponding to whether a given grid point is occupied by a particle or not. It is then unwrapped into a $1$-dimensional string in a way that preserves the spatial patterns or correlations to the extent possible when mapping a higher dimensional system onto a 1D string. Preserving correlations is important as the compression algorithms achieve compression by identifying repeating patterns within the string and replacing them with shorter descriptions. However, the drawback of this kind of discretization is that the results are grid-size dependent. At finer grid sizes, the compression is not very efficient, whereas at coarser grid sizes, there is information loss due to discretization errors. It is clearly less attractive to have an entropy estimate that depends on the choice of the level of discretization.

In this paper, we will consider CID. However, we propose a new method for measuring the information content of particulate systems in continuum space, which does not require the explicit discretization of the co-ordinates as discussed in previous studies. We then investigate the usefulness of CID as a quantitative measure of the entropy. We compare the CID measures with thermodynamic entropy estimates for simple systems in equilibrium, and then extend our studies to non-equilibrium systems of increasing complexity. We identify three classes of systems: (i) systems for which the CID measures show good agreement with other known entropy estimates, (ii) systems where the CID measures identify transitions but where the relationship to entropy is not clear and (iii) systems where the CID measures perform objectively worse than structure based entropy measures.

The rest of the paper is organized as follows. In Sec.~\ref{sec:methods}, we present the definitions and the different methods used to quantify the entropy. These include conventional entropy estimates \textit{viz.}, thermodynamic excess entropy and $n$-body entropy expansions, and compression based information measures. In Sec.~\ref{sec:results}, we present the details of the systems that we have studied and discuss our numerical results. Finally (Sec.~\ref{sec:summary}), we conclude the paper with a discussion of the implications of our findings. 

\section{Definitions and Methods}
\label{sec:methods}

\subsection{Thermodynamic Excess Entropy}

Consider a system in thermodynamic equilibrium having a number density $N/V\equiv\rho$ at a temperature $T$. The excess entropy $S_{ex}$ is defined as the difference between its total entropy $S$ and that of an ideal gas ($S_{id}$) at the same density and temperature:
\begin{equation}
S_{ex}(\rho, T) = S(\rho, T) - S_{id}(\rho, T).
\label{eq:excess-entropy}
\end{equation}

A commonly used technique for measuring the excess entropy in equilibrium simulations is thermodynamic integration~\cite{Vega2008}, where the excess entropy is obtained by integrating the equation of state along a reversible path from the ideal gas state to the target state:
\begin{equation}
\begin{split}
    \frac{S_{ex}(\rho, T)}{Nk_B} &= \frac{U_{ex}(\rho, T) - F_{ex}(\rho, T)}{Nk_BT}\\
    &= \frac{U_{ex}(\rho, T)}{Nk_BT} - \int_0^\rho \frac{p_{ex}}{k_BT\rho^2} \mathrm{d}\rho,
\label{eq:td-integration}
\end{split}
\end{equation}
where $U_{ex} = U - U_{id}$ is the excess internal energy, $F_{ex} = F - F_{id}$ the excess Helmholtz free energy and $p_{ex} = p - p_{id}$ the excess pressure.

\subsection{Pair Correlation Entropy}

The excess entropy can be expanded into an infinite series using Kirkwood's factorization~\cite{Kirkwood1942} of the $n$-particle distribution function~\cite{Green1952,Nettleton1958,Raveche1971,Wallace1987}:
\begin{equation}
S_{ex} = S_2 + S_3 + \cdots,
\end{equation}
where $S_n$ is the $n$-body contribution to the entropy. The pair correlation entropy $S_2$ usually makes the dominant contribution to $S_{ex}$, and is used as an approximate measure of $S_{ex}$ in studies of liquids and glasses. The ensemble invariant expression of $S_2$ in terms of the pair correlation function $g(\mathbf{r})$ is given as
\begin{equation}
S_2/Nk_B = -\frac{\rho}{2} \int_0^{\infty} \{g(\mathbf{r})\ln g(\mathbf{r}) - [g(\mathbf{r}) - 1] \} \mathrm{d}\mathbf{r}.
\label{eq:s2}
\end{equation}
For binary systems with two kinds of particles,
\begin{equation}
S_2 = \sum_{\alpha, \beta} x_{\alpha}x_{\beta} S_{2, \alpha\beta},
\label{eq:s2-partial}
\end{equation}
where $x_{\alpha}$ is the mole fraction of the $\alpha$ particle, and $S_{2, \alpha \beta}$ is the contribution from the partial pair correlation function $g_{\alpha \beta}(r)$ between $\alpha$ and $\beta$ particles. In the case of lattice systems, the integral in Eq.~\ref{eq:s2} is replaced by a sum over the discrete vectors $\mathbf{r}$.

\subsection{Computable Information Density}

Consider a string $X$ of alphabets of length $L$ whose information density we want to find. If $\mathcal{L}(X)$ is the binary code length of the shortest encoding that can reproduce $X$, then the computable information density (CID) of $X$ is defined as
\begin{equation}
I(X) = \frac{\mathcal{L}(X)}{L}
\label{eq:cid}
\end{equation}
If the string $X$ is compressed using the Lempel-Ziv algorithm (LZ77)~\cite{Ziv1977}, the binary code length $\mathcal{L}(X)$ is given as
\begin{equation}
\mathcal{L}(X) \leq C \log_2 C + 2 C \log_2(L / C),
\label{eq:lzlength}
\end{equation}
where $C$ is the number of distinct factors obtained by the LZ77 algorithm from the source sequence $X$~\cite{Martiniani2019}. % {Ziv78 appears to be a different algorithm.. Need to be careful.}.
But in order to compute the information density of an $n$-particle system, we must first represent it in a form that is suitable for processing by a compression algorithm. More importantly, this representation should preserve the information about the spatial ordering of the particles. We discuss below two such methods.

\subsubsection{Grid Based CID}
In the  approach of refs.~\cite{Martiniani2019,Avinery2017} the CID of an off-lattice $n$-particle system is computed by first dividing the system into a regular $D$-dimensional grid with a grid spacing $w$ such that a cell in the grid is occupied by at most one particle. Any grid $i$ which contains the center of a particle (or, for particles with a finite extent, is covered by a particle) is given an occupation number $c_i = 1$. All other grid cells are given $c_i = 0$. In other words, we convert the configuration of particle coordinates into a lattice of occupation numbers.
The $c_i$'s are then encoded into a single string by scanning the grid using a suitable space filling curve such as the sequential or spiral or Hilbert scan. Here, we use the Hilbert scan which gives the shortest average distance between two successive points along the scan. The Hilbert scan has the added requirement that the number of grid points should be a power of $2$. We then compress this string using the Lempel-Ziv algorithm to obtain the CID.

The CID thus computed is an estimate of the total information density or entropy of the system. In order to make comparisons with the thermodynamic excess entropy we need to subtract the corresponding CID of the ideal gas. To do this, we take a random configuration at the same density $\rho$ and calculate its information density $I_{id}$. The excess information density is defined as
\begin{equation}
I_{ex}(\rho) = I(\rho) - I_{id}(\rho).
\label{eq:excess-cid}
\end{equation}

An important point to note when computing the CID is that the length of the input string for the compression algorithm should be same across different particle densities in order for any data comparison to be valid. This is due to the fact that compression efficiency varies with the input string size. Therefore, for the grid-based CID, we effect changes to particle density $\rho$ by varying the number of particles $N$ and keeping the system size $L$ constant so as to keep the number of grid points constant. We then scale the resultant $I$ by $1/\rho$ so as to estimate the per particle information density.
Since the value of the grid-based CID depends on the grid size, any comparison with available thermodynamic entropy data will require the introduction of one scaling parameter, which we determine by least-squares fitting.

\subsubsection{Bit-Resolved or Scale-Free CID}

Introducing a grid-based discretization procedure for computing the CID of off-lattice systems, seems arbitrary and less attractive, as the CID can depend on the grid resolution. Below, we introduce a  method to convert a configuration of particles into a bit string without any discretization. To this end, we proceed as follows:
\begin{enumerate}
\item We use a Hilbert scan to order the particle coordinates. Note that the Hilbert scan has a resolution, but only to the extent that its resolution should be high enough to ensure that the ordering of the particles be unique. Any further increase in resolution makes no difference to the ordering of the particles. 
\item Normalize the particle co-ordinates to the range $[0,1]$, and represent them by $16$-bit integers, except in the case of the 1D gas of hard rods where some information is contained in bits 17-32.  The meaning of each bit is easily understood. For example, the leading bit of the $x$ co-ordinate tells whether the particle is in the left ($0$) or right ($1$) half of the box. Higher bits correspond to further sub-divisions of the box.
\item Take the value of the $j^\mathrm{th}$ bit of the particle co-ordinates to form the $j^\mathrm{th}$ binary string for compression. If $X_j^i$, $Y_j^i$ and $Z_j^i$ are the $j^\mathrm{th}$ bit of the co-ordinates of the particle $i$, then the input string is $X_j^1Y_j^1Z_j^1X_j^2Y_j^2Z_j^2X_j^3Y_j^3Z_j^3 \cdots X_j^NY_j^NZ_j^N$. 
\item Compress the binary strings using the LZ77 algorithm to estimate the compression density $I_j$ of $j^\mathrm{th}$ bit.
\begin{equation}
I_j \leq \frac{C_j \log_2 C_j + 2 C_j \log_2 (N / C_j)}{N},
\end{equation}
\end{enumerate}
where $C_j$ is the number of LZ77 factors for the $j^\mathrm{th}$ string. For all $j$, the $I_j$ thus computed cannot be larger than $I_R$, the compression density for a random string of $0$s and $1$s. However, for higher (i.e. less significant) bits,  $I_j$ approaches $I_R$. 
We now quantify the total information density as the area between the CID of the random string and that of the target system
\begin{gather}
%A =2^{I^{R}}D\ln(2)\sum_{j=1}^{nbits}\left(I^{R}-I_j\right)\Theta\left(\Delta_j\right) \notag \\
A =-\frac{D\ln 2}{I_R}\sum_{j=1}^{nbits}\left(I_{R}-I_j\right)\Theta\left(\Delta_j\right) \notag \\
\Delta_j =\frac{\left| I_j-\overline{I}_{R}\right|}{\sqrt{\overline{I_{R}^2}-\overline{I}_{R}^2}}-1.0
\label{eq:A_abs}
\end{gather}
% Here, we replace the three characters of each bit by an integer ranging from 0 to 9, keeping only one character on each bit. Thus $I_{R}$ is the CID of a random string of length $N$ with alphabet $\alpha=\{0..3\}$ for two-dimensional systems or $\{0..7\}$ for three-dimensional systems. 

The conversion factor $D\ln 2 / I_R$ is estimated based on considering the entropy of the ideal gas, which we summarise briefly. Let us consider first the simplest case of the $1D$ ideal gas. If the particles in the system are confined to the left half of the box, 
the entropy decrease -- compared to the case when the particles are uniformly distributed in the box -- is $\ln 2$ per particle. The bit strings representing the particle coordinates for the constrained system have $'0'$s in the first position,  with all other bits 
shifted to the right by one position. The corresponding $I_j$ curve thus shifts to the right by 1 bit, with a resultant increase in the (negative of the) area $A$ by  $I_R$. Thus the conversion factor to obtain the known change in entropy is $\ln 2/I_R$. In higher dimensions $D$, considering the scaling of coordinates in each of the $D$ dimensions by $1/2$ we obtain the conversion factor to be $D\ln 2/I_R$. The CID of any system approaches that of a random string at higher bits, which contains no information but is a source of numerical noise. Therefore we only consider the bits where $I < I_R$, as determined by the Heaviside function in the expression above. 

% {\color{magenta}[AB: My suggestion is to describe the above for 1d first, in which case the conversion factor would be $\ln 2/I_{R}$. And then mention that in arbitrary d, all the $D$ dimensions have to be scaled by $0.5$ so as to shift the curve by 1 bit to the right. So the conversion factor would be $d\ln 2/I_{R}$. $I_{R}$ is a simple and straightforward quantity to compute, rather than $\Delta A_{ig}$. The $\Theta$ function is difficult to understand. At the very least, some explanation should be there.]}

\begin{figure}
    \centering
    \includegraphics[width=0.5\linewidth]{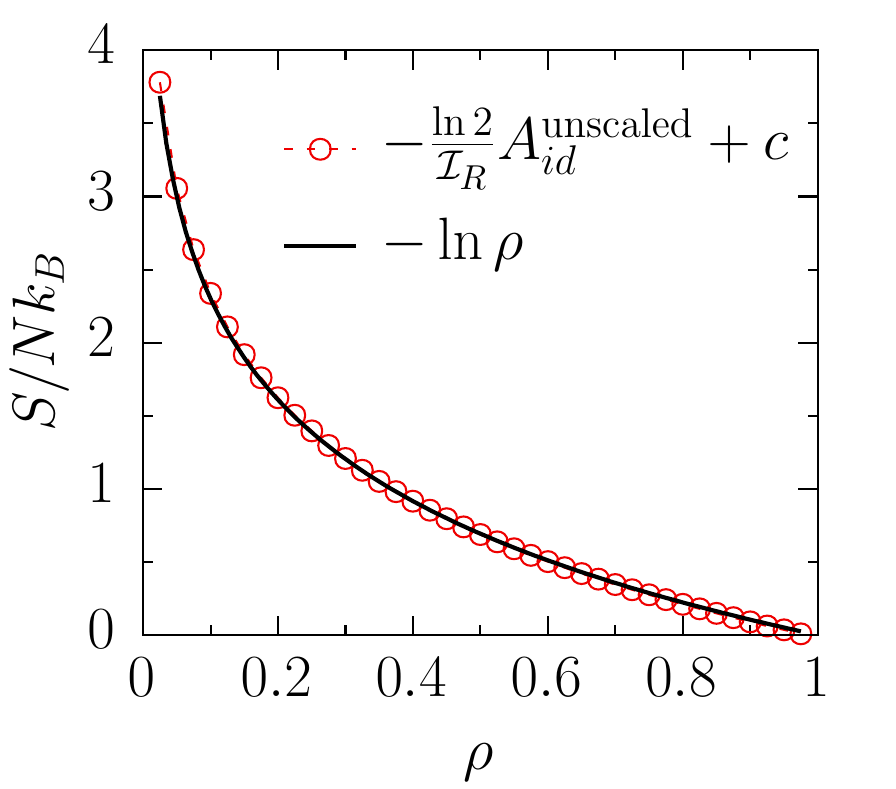}
    \caption{The total bit-wise information content of 1D ideal gas scaled by $-\ln 2 / I_{R}$, compared with the configurational part of the ideal gas entropy $-\ln \rho$. The constant shift $c = 8.838$.}
    \label{fig:A-scaling}
\end{figure}

We illustrate the validity of the above scaling in Fig.~\ref{fig:A-scaling}, where we have plotted the bit-wise information content of the 1D ideal gas scaled by $-\ln 2 / I_{R}$, as a function of density $\rho$. The constant shift $c$ is the only fitting parameter involved. Clearly, the scaled value of $A_{id}$ matches well with configurational part of the ideal gas entropy, $-\ln \rho$. For the rest of the results shown, we thus use the expression above for the bit-wise information density.

Note, however, that even an ideal gas $A$ is non zero, because adjacent ideal gas particles tend to have coordinates that differ only a little in the leading few bits. Hence, the meaningful way to quantify the excess CID of an $n$-particle system is to compute $A_{ex}$, the {\em difference} in the area $A_{id}$ for an $n$-particle ideal gas and the $A$ for $n$ particle system under consideration. Note that the expression for the scale-free excess entropy based on Eq.~(\ref{eq:A_abs}) contains no adjustable parameters and is therefore not subject to fitting.

An alternative (but not equivalent) way to compute the excess CID would be to perform a scale-free compression of the {\em distances} between successive particles in our Hilbert scan (and subtract the corresponding number for an ideal gas).  Below, we discuss both approaches.

%MZ: the prefix part of this equation is guessed as average information density

\section{Results and Discussion}
\label{sec:results}

We first briefly discuss two non-equilibrium model systems that have been studied in~\cite{Martiniani2019}
using  grid-based CID. We include these results only as a validation of our grid-based CID and as a test for the other entropy measures that we use in the remainder of this paper. \textmark{Note that for the non-equilibrium cases no unambiguous definition for the excess entropy exists. Hence, we compute $S_2$ as a placeholder for the excess entropy, and compare our CID estimates with it. For equilibrium systems, the difference between the excess entropy and $S_2$ is often small, making $S_2$ a credible approximation for $S_{ex}$. For the binary Lennard-Jones systems described below, available results show that $S_2$ indeed is a good approximation for $S_{ex}$~\cite{BanerjeeJCP2017}. Although inclusion of higher-order correlation entropies ($S_3$ {\it etc.}) may result in more accurate values for $S_{ex}$, we do not expect our results to change qualitatively.}

\subsection{The Manna model}

The Manna (``sand-pile'') model  is defined as follows: Initially, $N$ particles are randomly distributed on a $L \times L$ lattice, with each site $i$ having an occupation number $c_i$ which equals the number of particles at that site. Any site with $c_i$ greater than the threshold $z$ is active. At each step, all the particles on a randomly chosen active site are redistributed randomly over its neighbours. $N$ such steps constitute one cycle. As is well known, this model exhibits a non-equilibrium transition from an absorbing state to an active state as a function of the particle density. For densities below the critical density $\rho_c \approx 0.684$, the number of active sites decay to zero, {\it i.e.} the system reaches an absorbing state. For $\rho > \rho_c$, the system reaches a steady state with a well-defined fraction of active sites $f_a$. % Note that the Manna model is a lattice model, hence there is a natural grid and there is no point in using bit-by-bit data compression. % AB: I've commented this bit out temporarily, but we probably want to keep this and remove the second next paragraph.

\begin{figure}
\centering
\includegraphics[width=0.7\linewidth]{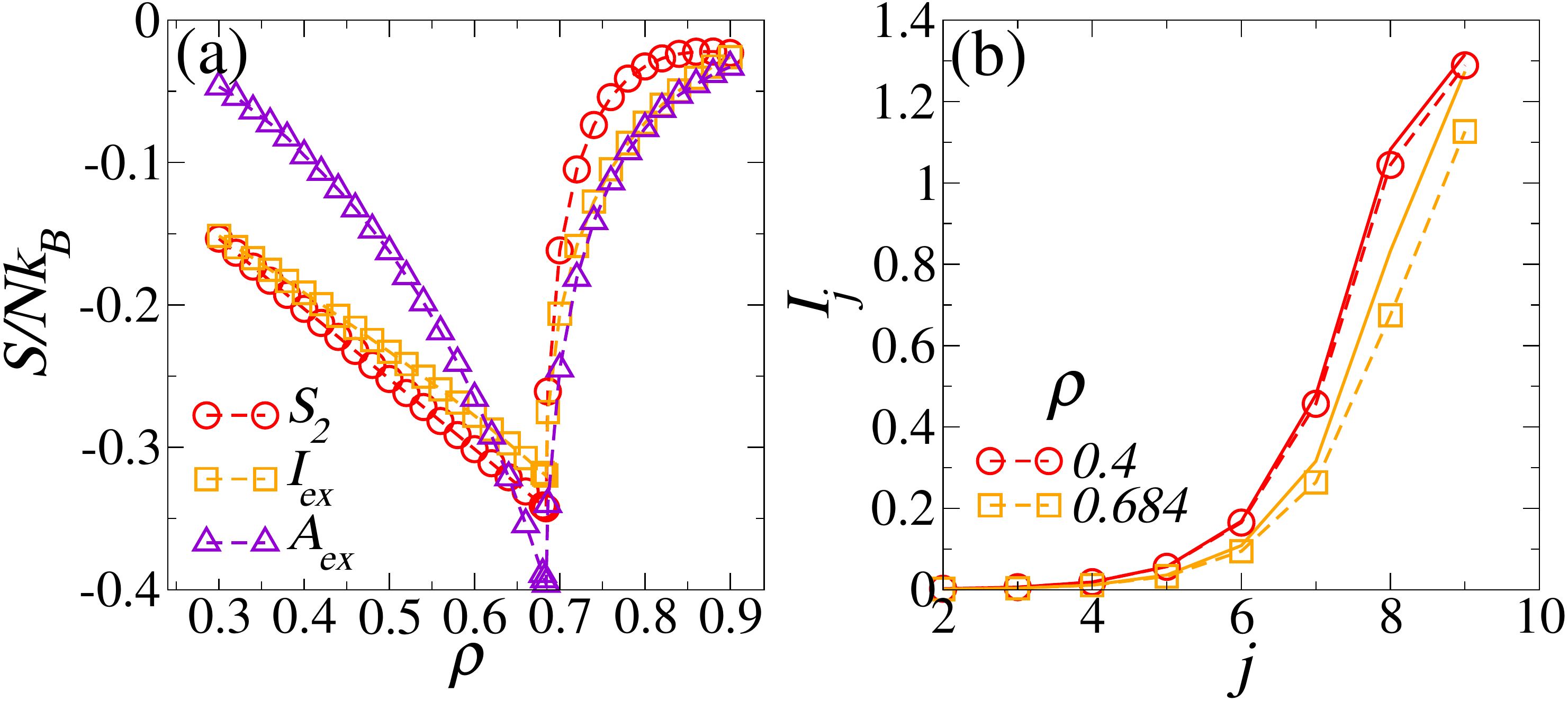}
\caption{(a) $S_2$, $I_{ex}$ and $A_{ex}$ as a function of $\rho$, for the Manna model on a $512\times 512$ lattice, in the steady state. The $I_{ex}$ data has been scaled to match $S_2$. (b) $I_j$ of the particle co-ordinates vs. $j$ for $\rho = 0.4$ and $0.684$. The solid lines are for the ideal gas at the corresponding densities.}
\label{fig:manna}
\end{figure}

To compute the CID, we use the Hilbert scan of the $c_i$s to generate the input string for data compression. In Fig.~\ref{fig:manna}, we show the pair correlation entropy $S_2$, the excess CID $I_{ex}$, and the excess area of the bit-wise CID $A_{ex}$ for the Manna model on a $512 \times 512$ lattice in the steady state ($2 \times 10^5$ steps), with the threshold for activity $z = 1$. The data has been averaged over $128$ samples. {Note that the $S_2$ shown here is \textit{per particle}, not per lattice site.} The $I_{ex}$ data has been scaled by a factor of $0.245$, to match the $S_2$ data. We see that both $A_{ex}$ and $I_{ex}$ show a minimum at the critical density $\rho_c$. {Note that, to facilitate comparison with $S_2$, we have subtracted the CID corresponding to a random distribution of particles over cells}.

In Fig.~\ref{fig:manna}(b), we show $I_j$ of the particle co-ordinates as a function of the bit depth $j$ for two densities $\rho = 0.4$ and $0.684$. The ideal gas reference is also taken on a lattice. The corresponding curves for the ideal gas reference are shown by solid lines. As the points are on a $512^2$ lattice, all the bits above $j = 9$ are redundant and carry no information. We find that all three entropy measures are qualitatively similar.

\subsection{The Random Organization Model}
The random organization model (ROM) is a continuum analog of the Manna model. It was first introduced to study the reversible state to irreversible state transitions observed in periodically sheared colloidal suspensions~\cite{Corte2008}.  Here we study {a simple static} variant of the model which is defined as follows: $N$ discs each of diameter $d$ are initially randomly distributed over a $L \times L$ square box. Any overlapping discs are considered active. At each step, \textit{all} active discs are given a small displacement in a random direction. The random displacement imparted to the particle at each step was $d/3$. The state of the system is determined by the packing fraction $\phi = N a_0 / A$, where $a_0$ is the area of a disc and $A = L^2$. System undergoes a dynamical phase transition (from absorbing to active), as $\phi$ is increased to a critical density $\phi_c \simeq 0.362$.  

\begin{figure}
\centering
\includegraphics[width=0.7\linewidth]{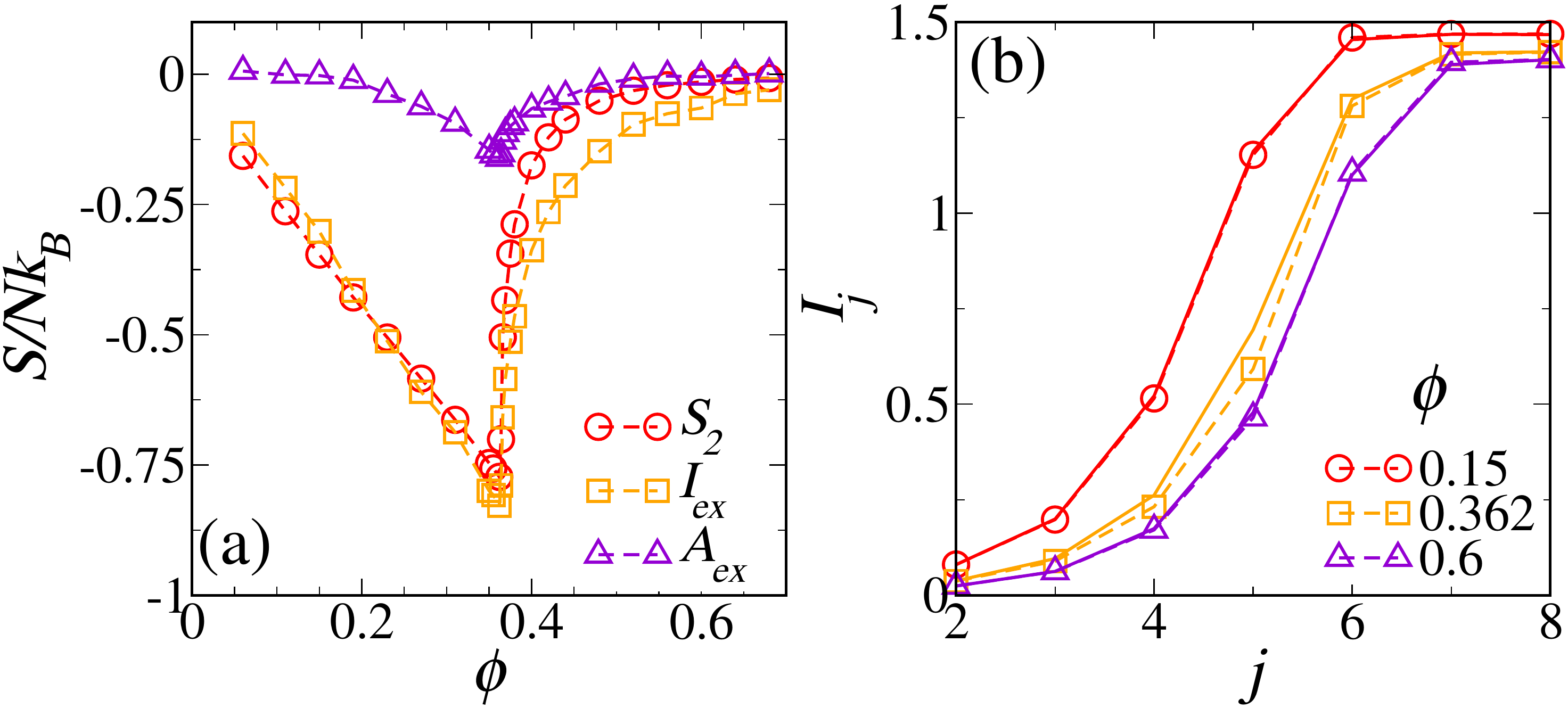}
\caption{(a) Various entropy estimates for the random organization model as a function of $\phi$. The grid-based CID is scaled to match the $S_2$ data. The critical packing fraction for the absorbing-to-active state transition is $\phi_c \approx 0.362$. (b) The CID of the particle co-ordinates $I_j$ as a function of bit depth $j$. The dashed lines correspond to ROM and the ideal gas curves at the corresponding density are shown by solid lines.}
\label{fig:rom}
\end{figure}

We control the volume fraction of the system by varying the number of particles in a box with a fixed size $L = 100d$. The configuration has been digitized on a $256 \times 256$ grid, leading to a grid width of $\sim 0.4d$. The data that we show correspond to the steady state (after $2 \times 10^5$ steps) and has been averaged over $64$ samples. Fig.~\ref{fig:rom} shows the $S_2$, the grid based CID $I_{ex}$ and the scale-free CID $A_{ex}$ as a function of $\phi$ for the ROM. The grid-based CID data is scaled by a factor of $6$ to match the $S_2$ data. Similar to the Manna model, $S_2$ and $I_{ex}$ show a sharp (cusp-like) minimum at $\phi \approx 0.362$. The excess information $A_{ex}$ computed from the bit-wise CID of particle co-ordinates also shows a cusp-like minimum near the transition, even if it does not show quantitative agreement with the other two entropy measures.

\subsection{1D Hard Rod Gas}
We now consider a number of off-lattice equilibrium systems, for which the entropy is well defined. 
 this allows us to make a quantitative, rather than just qualitative comparison between the entropy of a system and the estimate based on bit-wise data compression.
As a first test, we consider a $1$-dimensional gas of hard rods as the entropy and radial distribution function of this off-lattice system are known analytically. The equation of state of the hard-rod gas was first derived by Tonks in 1936~\cite{Tonks1936}. From the analytical equation of state, we then obtain an analytical expression for the excess entropy $S_{ex}$. For a system with $N$ rods of size $\sigma$ in a box of length $L$, the excess entropy has a simple form:
\begin{equation}
S_{ex}/Nk_B = \ln(1 - \rho \sigma),
\label{eq:tonks-excess}
\end{equation}
where $\rho = N / L$ denotes the number density. In addition to the exact excess entropy, we can obtain $S_2$ by inserting into Eqn.~\ref{eq:s2} the analytical expression for the pair correlation function:
%\begin{multline}
\begin{equation}
g(r) = \sum_{n=1}^{\infty} \frac{\Theta(r-n)}{1-\rho \sigma}\left(\frac{\rho \sigma}{1-\rho \sigma}(r-n)\right)^{n-1} \times \frac{1}{(n-1)!}\exp\left(-\frac{\rho \sigma}{1-\rho \sigma}(r-n)\right),
\label{eq:tonks-gr}
\end{equation}
%\end{multline}
where $\Theta$ is the Heaviside step function and $r$ is measured in units of $\sigma$.

For computing the CID, we generate the configurations through Monte Carlo sampling as follows. The system is initialized in a state, with all $N$ rods at a separation of $1/\rho$ along the length $L$ of the box. We then carry out standard Monte Carlo equilibration for $10^5$ Monte Carlo steps (MCS) (one MCS equals $N$ single-particle trial moves). The equilibrium properties were then sampled during the subsequent $10^5$ MCS, at intervals of $100$ MCS, giving a total of $10^3$ samples. We perform density changes by varying $N$ and keeping $L = 10^5 \sigma$.

\begin{figure*}
\centering
\includegraphics[width=\linewidth]{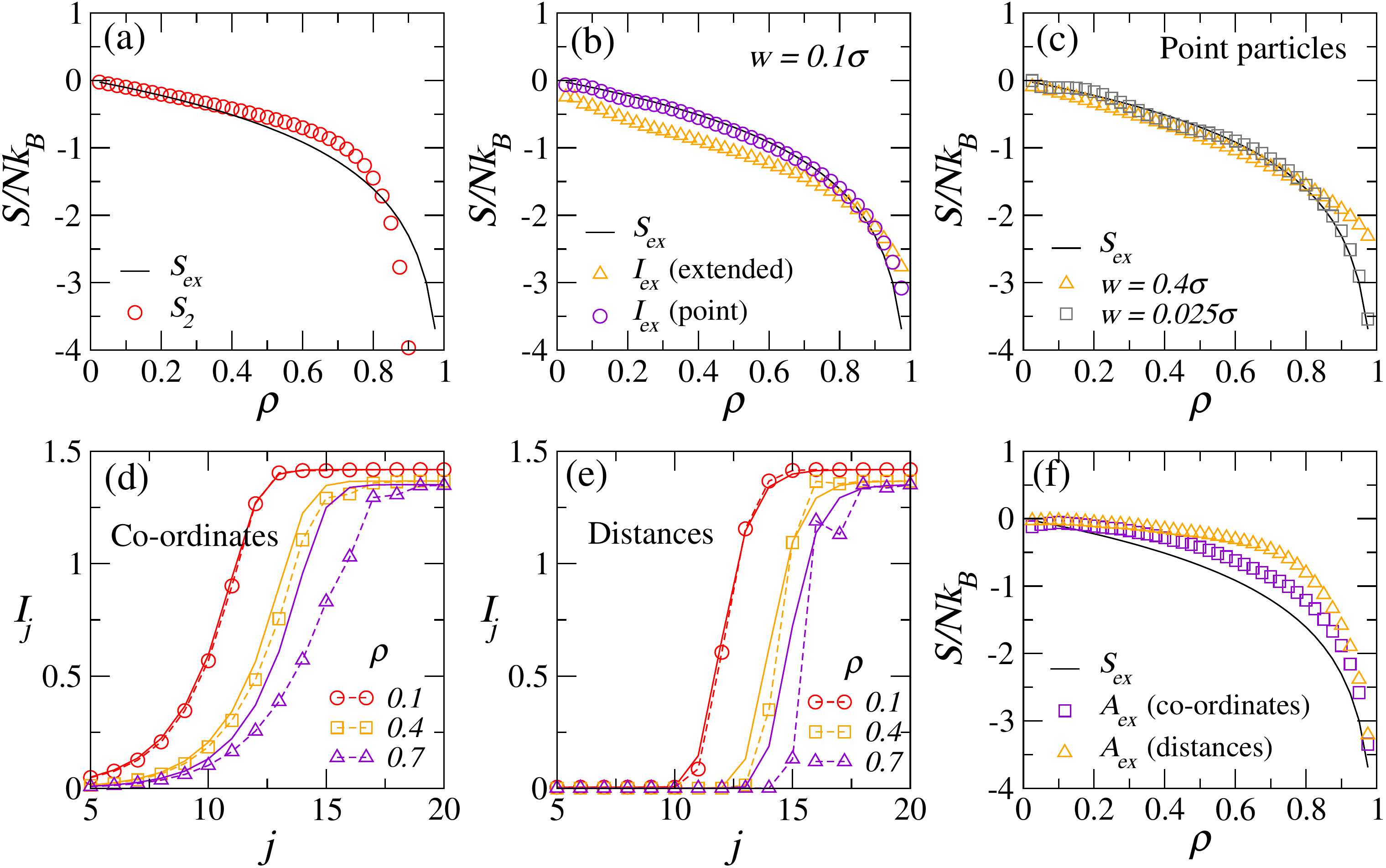}
\caption{Plots of various entropy estimates for the $1D$ hard rod gas: (a) $S_2$ and $S_{ex}$ vs. $\rho$. (b) Comparison of grid-based CID estimates with $S_{ex}$. The two data sets are when the discretization is done by taking the rods to be points or extended particles. (c) Variation of CID with grid size used for discretization, for the point representation. The grid-based CID data have been scaled (not shifted) to match $S_{ex}$. (d) Bit-wise CID of the particle co-ordinates as a function of the bit-depth $j$. (e) Bit-wise CID of successive particle distances as a function of the bit-depth $j$. The dashed lines correspond to the hard-rod system at the given density, and the solid lines correspond to the ideal gas at the same density. (f) Excess information density $A_{ex}$ for the bit-resolved CIDs.}
\label{fig:tonks}
\end{figure*}
In Fig.~\ref{fig:tonks}(a), we show $S_2$ and $S_{ex}$ vs. $\rho$ for the $1D$ hard rods. The $S_2$ values are calculated from $g(r)$ obtained from numerical simulations, and have been verified to match with the values computed using the analytical expression of $g(r)$ given in Eq.(\ref{eq:tonks-gr}). % \DFcomment{Before, you suggested that you used the analytical $g(r)$} % AB: The text has been updated.\\
For densities $\rho \lesssim 0.82$, $S_2$ underestimates $S_{ex}$, whereas above it overestimates $S_{ex}$. {This is also generally found to be the case for simple liquids where, for densities below (above) the freezing point $S_2$ underestimates (overestimates) $S_{ex}$}. We now turn our attention to the various compression-based information density estimates. As discussed in the previous section, we have scaled the CID to obtain an optimal match with the $S_{ex}$ data. Fig.~\ref{fig:tonks}(b) shows the grid-based CID measures. The two data sets correspond to the two different discretization procedures (point vs. extended particles) used to convert the particle co-ordinates into a grid of occupation numbers (see Sec.~\ref{sec:methods} A.1). A grid spacing of $w = 0.1 \sigma$ was used for the discretization of the simulation box. We find that the point representation matches well with the $S_{ex}$ data except at high densities. Taking the finite extent of the rods into account, yields a worse CID estimate of $S_{ex}$. For the point representation, a grid size of $\approx \sigma / 10$ gives the best results % \DFcomment{Those results are not shown}. % AB: The data is now shown for $w=0.1\sigma$
Smaller grid spacing leads to lesser agreement with the excess entropy data (see Fig.~\ref{fig:tonks}(c)).

% AB: The scale factor for matching with $S_{ex}$ increases with decreasing grid size for both the finite as well as the point particle representations. It roughly follows: 1/sf \propto log(gs), where sf and gs are the scale factor and grid size respectively. This is not the most perfect scaling (the data still curves slightly on a linear-log plot), and has been checked only for the point representation.

Next, we show the bit-wise information density of the particle co-ordinates and the inter-particle distances as a function of the bit index $j$ in Fig.~\ref{fig:tonks}(d) and (e), respectively. The dashed lines correspond to the hard-rod system at a given density, and the solid lines correspond to the ideal gas at the same density. $j = 1$ represents the most significant bit (MSB). The $I_j$ curves exhibit the following typical behaviour: $I_j$ starts with a low value for $j=1$, increases gradually with $j$ and saturates to a maximum after a certain bit depth. The maximum value of $I_j$ corresponds to a fully random (and hence incompressible) string. Because of the sorting of the particles, the strings formed by the lower bits tends to be correlated and hence have lower CID. For the CID of the particle co-ordinates, the ideal gas has a higher information density than the hard rod gas at any density. This is to be expected as the ideal gas particles are completely uncorrelated.
% However, for the CID of the particle distances, the ideal gas has the lower information density. This finding suggests that the CID excess entropy based on particle distances is unphysical. % AB: This is no longer the case. The curves have been updated. I was normalizing the particle distances by the maximum of their distribution. Using the box-length to normalize the distances gives the present curves.
This is also generally true for the CID of the particle distances. However, for $\rho \gtrsim 0.6$, the $I$ vs. $j$ curve shows a drop in $I$ at $j \approx 16$. This is because the distribution of inter-particle distances get narrower with increasing density 
%(see Fig.~\ref{fig:s4}). 
As a result, the inter-particle distances are highly correlated, leading to a drop in the information entropy.
The corresponding area difference $A_{ex}$ is shown in Fig.~\ref{fig:tonks}(f). We note that even though the position-based  bit-wise CID does not agree quantitatively with the thermodynamic entropy, the difference is not very large, and might be due to finite-size effects.
We see that the for the CID of the co-ordinates, the excess information content matches well with the excess entropy, whereas that of the inter-particle distances does not. In what follows, we will not consider the finite-particle representation for the grid-based CID or the distance-based bit-wise CID estimates.
 
\subsection{Two-dimensional melting for soft-core systems}

\begin{figure*}
\centering
\includegraphics[width=\linewidth]{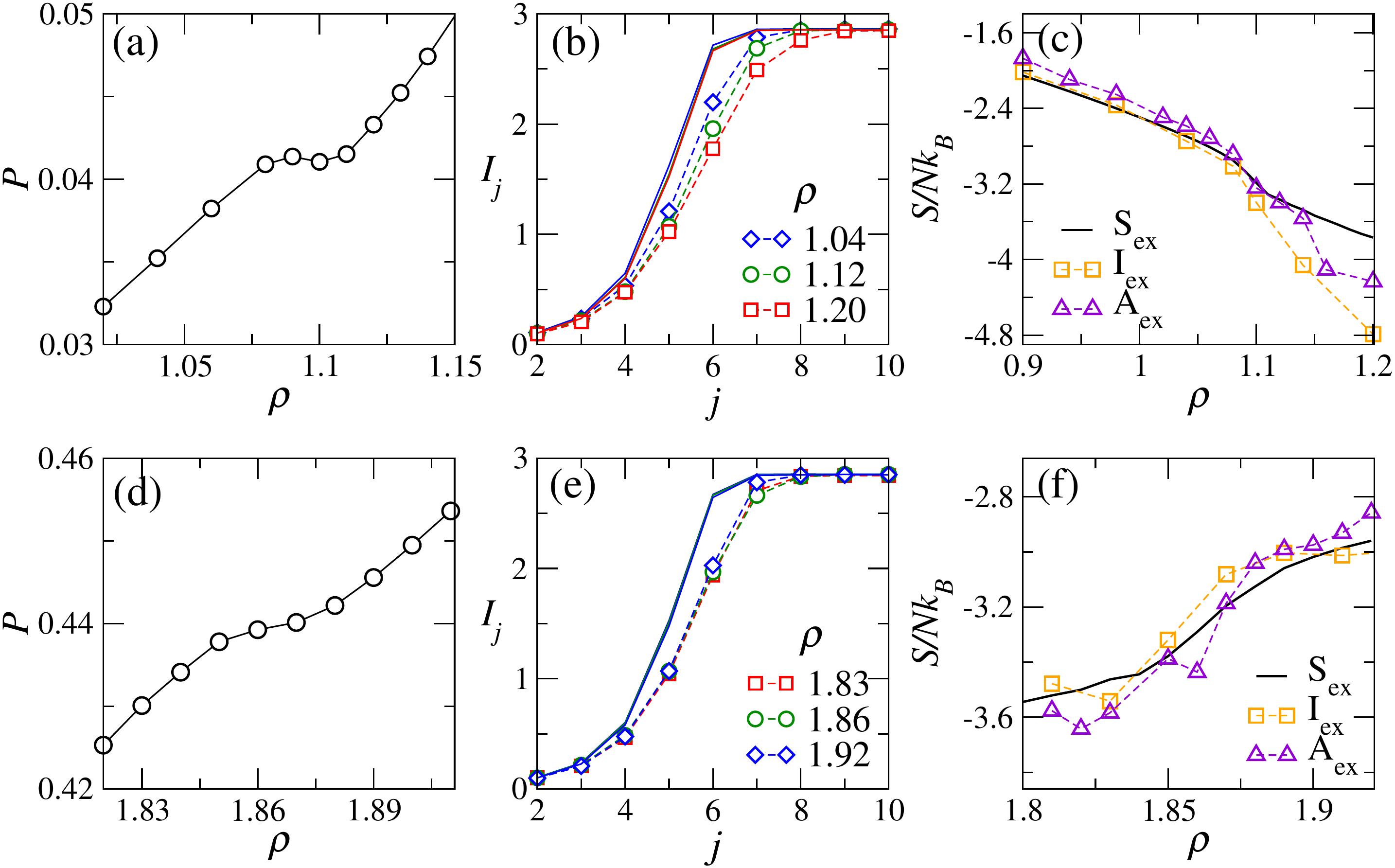}
\caption{\label{fig:mfig1} The isothermal equation of states at $T=3.3\times 10^{-3}$ for harmonic systems in a wide range of density. (a) At low density, the equation of states shows Mayer-Wood loop. (d) At high density, the pressure increases monotonically.  The compression information density as a function of bit-depth $j$ at (b) low and (e) high densities. {The dashed lines correspond to the soft-core systems at different densities, and the solid lines correspond to ideal gas systems at same densities. The rhombus-shaped symbols denote liquid states, the circles are hexatic states and the squares are solids.} (c) and (f) The various excess entropy $S_{ex}$, $I_{ex}$ and $A_{ex}$ as a function of density. The curves of $I_{ex}$ are scaled by multiplying a fitting parameter $215$ and $195$ respectively, and the curves of $A_{ex}$ is scaled with fitting parameter $2.2$ for (c) and (f).}
\end{figure*}

Another equilibrium system that we considered is the two-dimensional mono-disperse soft-core system exhibiting re-entrant crystallization with a maximum melting temperature $T_m$ at  density $\rho_m$. In this system we can separate the effect of the density and ordering on the entropy. In fact,  this system undergoes a liquid-hexatic transition that, depending on density, may be either first order  ($\rho < \rho_m$) or continuous (at higher densities)~\cite{Zu2016}. We consider a soft-disk model system, where pair particles interact via finite range and purely repulsive potential
\begin{equation}\label{eq:Harmonic}
V(r_{ij})=\frac{\epsilon}{\alpha}\left(1-r_{ij}/\sigma\right)^{\alpha}{\Theta}\left(1-r_{ij}/\sigma\right)
\end{equation}
where $r_{ij}$ is the distance between particle $i$ and $j$, and $\Theta$ denotes the Heaviside function. The potential vanishes continuously at  $r_{ij}=\sigma$, where $\sigma$ is the diameter of the disk. The exponent $\alpha$ determines the softness of the potential.  Here, we choose $\alpha = 2.0$ (the ``harmonic'' model).  $\epsilon$ and $\sigma$ define our units of energy and length respectively. 
The maximum melting temperature $T_m$ for the harmonic system is approximate $7.10 \times 10^{-3}$ at a crossover density $\rho_m\approx 1.42$. 

{We consider a square simulation box of sides $L = 58\sigma$, and vary $\rho$ by changing the number of disks $N$.} We slowly quench the high temperature equilibrium liquid states until target temperature $T=3.3 \times 10^{-3}$. Then we equilibrate the system at constant volume and temperature for at least $10^7$ MD steps and configurations are sampled for next $10^7$ MD steps at intervals of $10^5$ MD steps. Simulations were performed both below and above $\rho_m$, using a parallelized version of LAMMPS~\cite{LAMMPS} . 

In Fig~\ref{fig:mfig1} $(a)$, the isothermal equation of states $P(\rho)$ shows a Mayer-Wood loop~\cite{Mayer1965} at low density, suggesting first-order coexistence of the isotropic liquid and hexatic phases. By fitting the curve with a tenth-order polynomial and determining the boundaries of coexistence with a Maxwell construction, the coexistence region is found to be located in the density range  $\rho$=1.08 to 1.11. 
When we increase the pressure by a factor of around 10, we find a hexatic to liquid transition at $\rho\approx 1.865$ (see Fig~\ref{fig:mfig1} $(d)$). As both the low-density and the high-density phase transitions are effectively free of hysteresis, we can compute  the thermodynamic excess entropy from the  equation of state by thermodynamic integration (see Fig~\ref{fig:mfig1} $(c)(f)$).

For the grid-based CID, we discretize the configuration with a square grid fine enough that no more than one disk can occupy a grid cell, resulting in a total number of cells equaling $2^9 \times 2^9$, and a bin-size of approximately $\sigma/8$.
% We then scan each cell along the Hilbert curve to generate a compressed string with size $2^8 \times 2^8$, in which 1 indicates occupied cell and 0 indicates empty cell. For the scale-free CID, we choose same Hilbert scan size to order disks. The much larger \DFcomment{Than what?} Hilbert scan size does not change the order of disks. As before we convert the coordinates of the particles to 16-bit integers. We thus we obtain two data sets: $\{x_j^i\}$ and $\{y_j^i\}$, where ${i=1,...N, j=1,...,16}$. The compressed string $X_j$ is composed of $x$- and $y$-value at each bit for all disks, $X_j=\{x_j^iy_j^i, i=1,...N\}$. % AB: Commented out because much of this is already there in the methods section.
In Fig.~\ref{fig:mfig1} $(b)(e)$, we plot the compression density as a function of the bit-depth $j$ for target systems and ideal gas systems at same density with the alphabet $\alpha=\{0,1,2,3\}$ indicating the combination of x and y coordinates. As expected, at lower bits the system has less information, because the box is divided into four parts and only 4 string patterns form the compressed string.
For higher (less significant) bits, the compression density approaches that of a random string, indicating that, beyond these bits the coordinates provide no information on the physical state of the system. Here, we just consider the value at lower bits where the compression density is less than that of random string. For more ordered structures, the bit-depth at which the CID approaches that of the random string is higher. As a result, the value of $I_j$ of solid states is less than that of liquid states at same bit-depth $j$.

In Fig.~\ref{fig:mfig1} $(c)(f)$, we compare the excess entropy obtained by the three methods mentioned above. We note that the absolute value of the bit-wise CID entropy differs significantly from the thermodynamic  entropy. However, the qualitative density dependence is similar. This is best seen by scaling  the CID results with an adjustable  fitting parameter. We stress that the fact that a scale adjustment is necessary, seriously undermines the role of the CID as a quantitative entropy estimator.  Yet,  we note that, after this {\em ad hoc} rescaling, all entropies behave qualitatively similarly, although the two CID entropies differ from the thermodynamic $S_{ex}$ and from each other. Importantly, at higher densities, all estimates of the entropy increase with density, demonstrating that all method are more sensitive to positional correlations than to density. 

\subsection{Jamming transition for bi-disperse systems in 2D}

Moving from equilibrium to non-equilibrium off-lattice models, we next consider jamming in a bi-disperse two-dimensional system.  A system is said to be jammed  when it develops a yield stress~\cite{Hern2003}. Jamming transitions have been observed in a variety of systems, such as granular systems, colloids and cells~\cite{Zhang2009,Bi2011,Bi2016}.  Here we study various entropy measures for a jammed system. Such systems are not in an equilibrium state, yet are not driven either.

{Here, we consider a two-dimensional equimolar binary mixture of harmonic soft disks ($\alpha = 2$ in Eq.~\ref{eq:Harmonic}) in a square box of sides $L=150\sigma_S$, where $\sigma_S$ is the width of the smaller disk, at zero temperature.} The ratio of diameters of big to small disks is $1.4$, so as to suppress crystallization. The systems are prepared by quenching from initially random states. Here, we use a fast inertial relaxation engine (FIRE) algorithm~\cite{Bitzek2006} to implement the minimization of the potential energy. Periodic boundary conditions are applied. The simulations were performed on an ensemble of 1000 harmonic systems over a wide range of volume fractions. 

\begin{figure}
\centering
\includegraphics[width=0.7\linewidth]{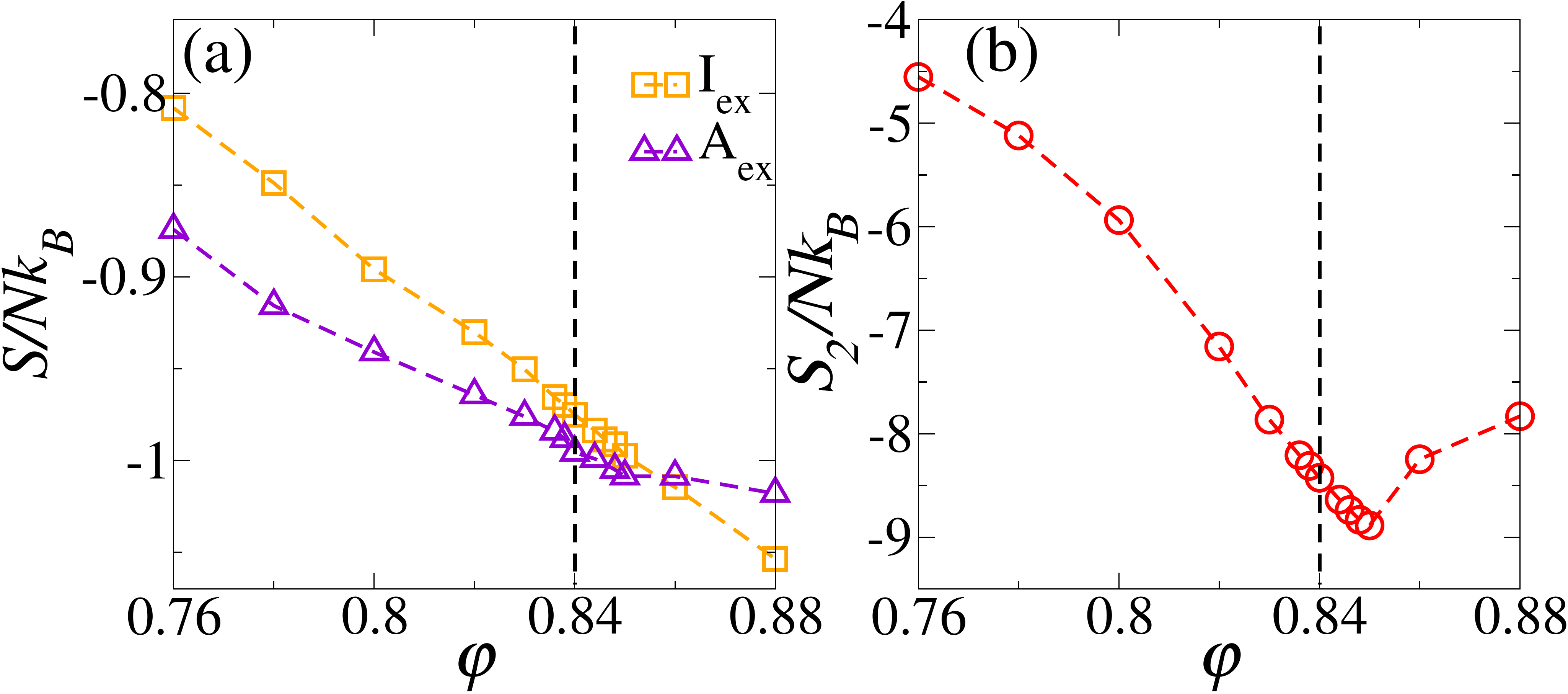}
\caption{\label{fig:mfig3} The excess compression information density $I_{ex}$ and $A_{ex}$ decreases smoothly with volume fraction increases. The curve of excess grid based CID $I_{ex}$ is scaled by a factor of 2.5. The dashed line indicates the location of point J. {(b) Pair correlation entropy $S_2$ as a function of the volume fraction.}}
%(b)The excess CID $A_{ex}$ as a function of system size $N$ at $\phi=0.88$ and $0.80$. The dashed lines correspond to linear fits for the 3 largest sizes fro each $\phi$.}
\end{figure}

The configurations were discretized with a grid size approximately $\sigma_S/4.0$ and the positions of the disks are sorted using a $512 \times 512$ Hilbert scan for bit-wise CID calculation. We convert the configurations into binary strings ignoring the type of disks, \textit{i.e.}, we only consider if a grid cell is occupied or not. % We used both the grid-based and the bit-by-bit compression resolution.
Fig~\ref{fig:mfig3}(a) shows the excess compression density $I_{ex}$ and $A_{ex}$ obtained from the bit-wise compression of the particle co-ordinates, as a function of the volume faction. Both $I_{ex}$ and $A_{ex}$ decrease smoothly as the volume fraction increases across the jamming point $\phi_J \approx 0.84$. The curve of $I_{ex}$ is scaled to match the values of bit-wise CID. {In Fig.~\ref{fig:mfig3}(b), we show the pair correlation entropy $S_2$ as a function of the packing fraction $\phi$. While the first peak of $g(r)$ is supposed to diverge at the jamming point $J$ due to the formation of contacts, when numerically computing the correlations, this divergence is not captured due to the finite width of the $g(r)$ bins. The $S_2$ results shown here must therefore be interpreted accordingly. We find that $S_2$ shows a minimum near the jamming point $J$. The CID measures however do not show this feature.}

%\textcolor{red}{\DFcomment{I do not understand this sentence.} The difference between compression density and entropy of source approaches 0 as $N \rightarrow \infty$ by using LZ77 algorithm~\cite{Wyner1997,Savari1998}. While the convergence rate for lots of positive entropy sources is slow, as shown in Fig~\ref{fig:mfig3}(b), the excess compression density $A_{ex}$ converge to the extrapolated values $A_{ex}^{\infty}$ as $\sim log_2log_2(N)/log_2(N)$. The results for three-dimensional lennard-Jones fluid would give further discussion in which case the thermodynamic entropy is a feasible.} \DFcomment{This definition of extensive is new to me}. \DFcomment{This statement is unclear - anyway, the results deserved further discussion

\subsection{Three-dimensional Lennard Jones fluids}
\begin{figure}
\centering
\includegraphics[width=0.7\linewidth]{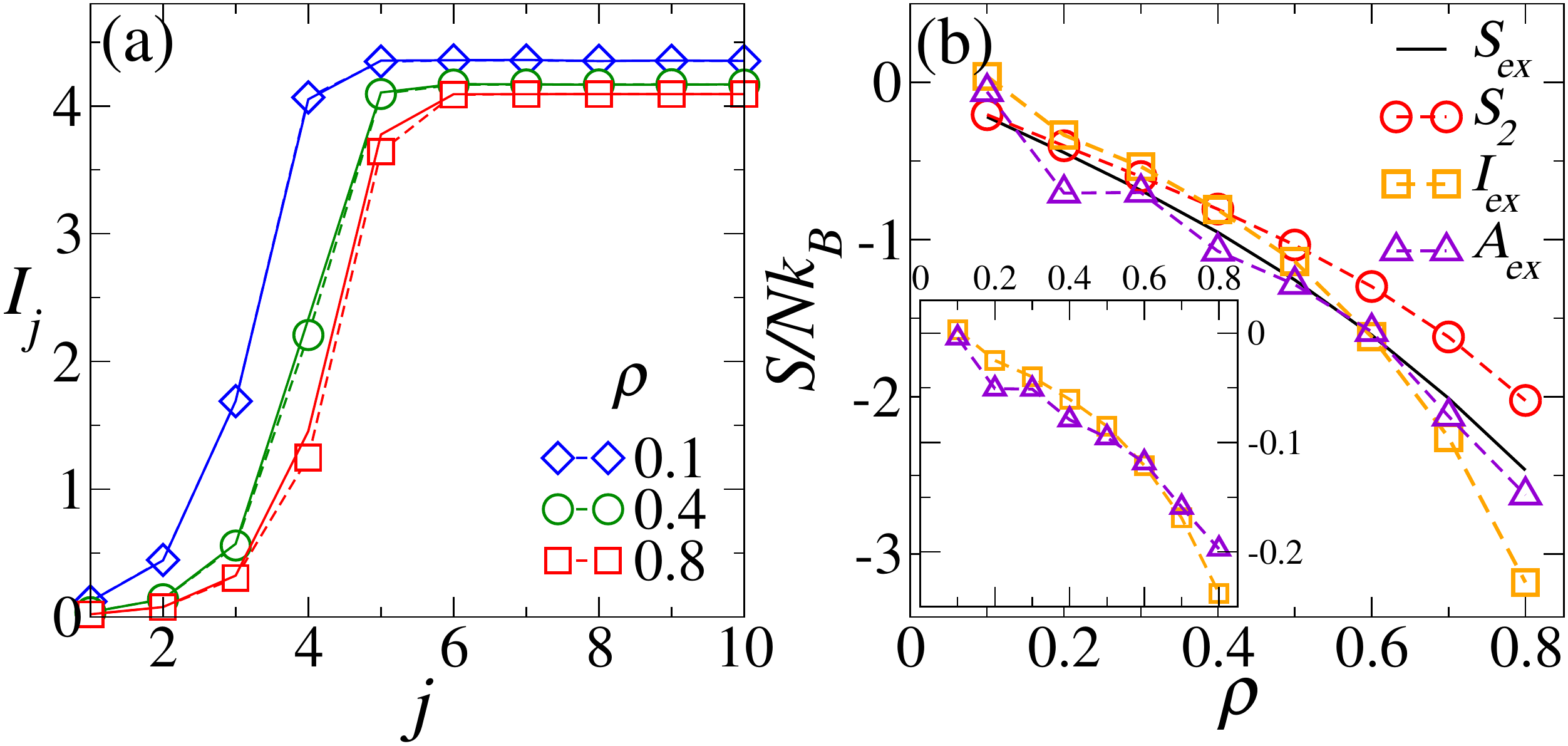}
\caption{\label{fig:mfig2} (a) The CID as a function of bit-depth for three-dimensional Lenard Jones systems at various densities. {The dashed curves correspond to the target system at the given density, and the solid lines denote the ideal gas at the corresponding density.} (b) The various excess entropy as a function of density at $T=2.0$. The curve of grid-based CID scaled by a factor of $8.53$. {The inset shows the values of $I_{ex}$ and $A_{ex}$ as a function of density calculated via Eq.~(\ref{eq:A_abs}).}}
\end{figure}

Little work has been done on data-compression based entropy estimates in dimensions higher than two (see, however, ref.~\cite{Avinery2017}). As a test of the various entropy estimators, we study several three-dimensional systems, both in  and out of equilibrium. 
First, we consider a three-dimensional Lennard-Jones fluid to test the generalization of bit-resolved CID in higher dimensions. {We performed constant-volume and constant-temperature molecular dynamics simulations on a system of Lennard-Jones particles without shifted potential in a cubic box of sides $L = 29\sigma$}. The potential cutoff is $4.0\sigma$, where $\sigma$ is the Lennard-Jones diameter. From the simulations, we obtain the equation of state at $T=2.0$ in the density range from $0.1$ to $0.8$. The excess thermodynamic entropy calculated by thermodynamic integration is consistent with previous works as shown in Fig.~\ref{fig:mfig2} (b)~\cite{Johnson1993,Frenkel-Smit}. 
 
We order the coordinates of all particles in a cubic 3D system using a $128 \times 128 \times 128$ three-dimensional Hilbert scan curve. The configurations are sampled at intervals of $10^5$ MD steps, giving a total of $10^3$ samples starting from ten independent equilibrium liquid states. By performing bit-wise data compression on the coordinates of the particles, where the $j$-th bit of the $x$, $y$ and $z$ coordinates were combined into a single bit string of length $N$ with an alphabet $\alpha=\{0,1,...,7\}$.

We also calculate the structural entropy $S_2$ and grid-based CID $I_{ex}$. Here, we discretize the configuration on a $128\times 128\times 128$ cubic lattice, where the grid size is small enough that only one particle occupies a grid cell. 
As shown in Fig.~\ref{fig:mfig2} (a), the differences of compressed information density of each bit between ideal gas and fluids are small, especially at low density.  As shown in Fig.~\ref{fig:mfig2} (b), the values of $I_{ex}$ show a large deviation from the thermodynamic excess entropy at higher densities.  Similarly, the bit-wise CID $A_{ex}$ can only be made to resemble the thermodynamic entropy scaling with an {\em ad hoc} factor.  The inset shows the results calculated from Eq.(\ref{eq:A_abs}) without rescaling. Clearly, the excess entropies obtained  from data compression without adjustment are much smaller than the values obtained by thermodynamic integration. In fact, the difference in scale is close to two orders of magnitude. It is conceivable that part of this very large discrepancy is due to finite size effects in the value of the CID, but this is bad news too, as the finite size correction to the CID appears to converge rather slowly with system size, suggesting that the extrapolation to the thermodynamic limit (if this is indeed  the significant reason for the quantitative difference) may not be feasible. 
%(see Supplementary Information). However, that is bad news too: as the finite-size correction to the CID converge extremely slowly ($\sim (\ln\ln N)/\ln N$), the extrapolation to the thermodynamic limit is, in practice, not feasible:  in other words: the CID-based entropy of a $3D$ off-lattice system can, in general, not reproduce the thermodynamic entropy. 
Both the structural entropy $S_2$, grid based $I_{ex}$ and bit-wise $A_{ex}$ CID  decrease as density increases. Not surprisingly, the structural entropy $S_2$ agrees well with the total excess entropy at low densities.

\begin{figure}
\centering
\includegraphics[width=0.45\linewidth]{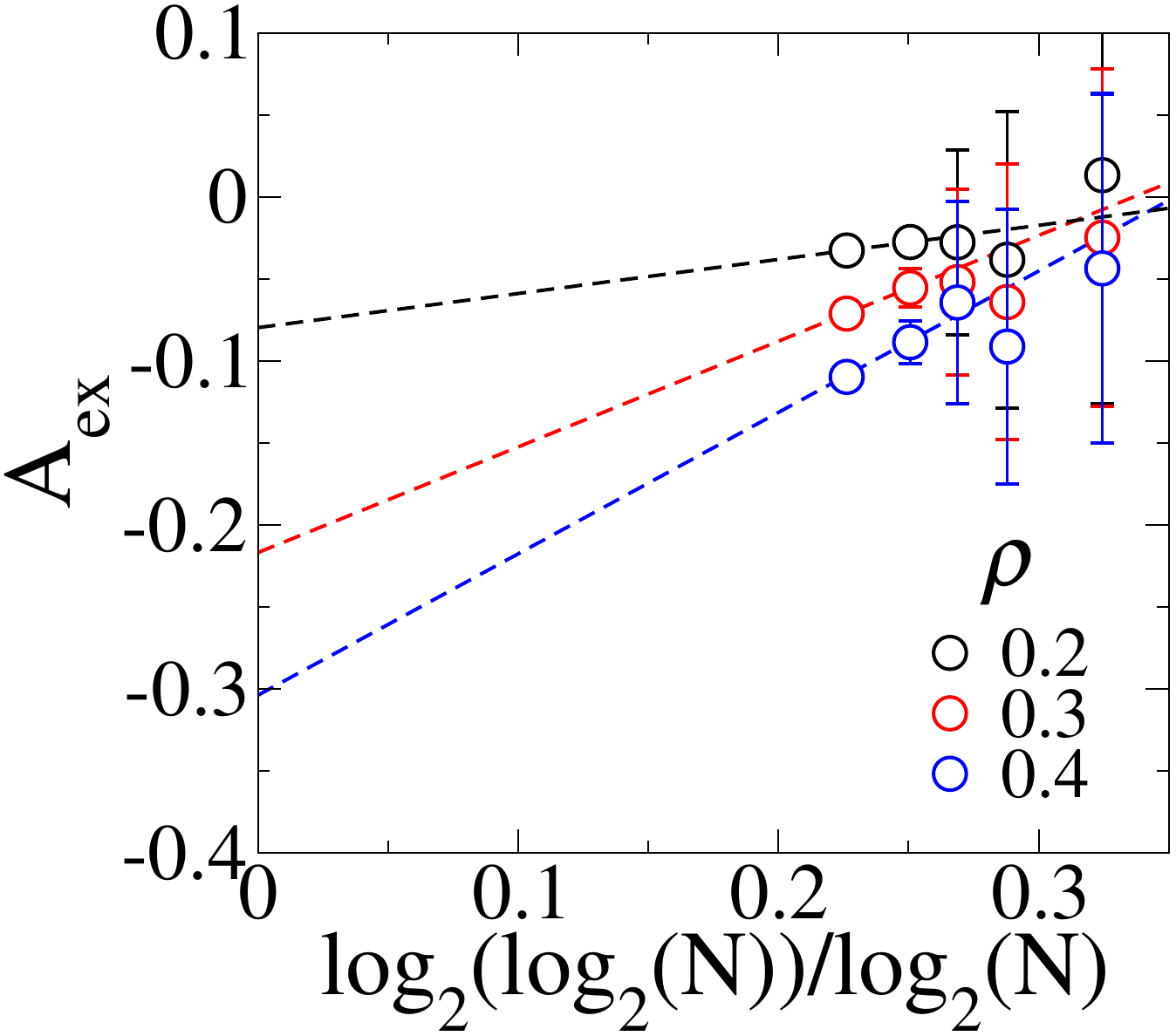}
\caption{\label{fig:finite_size} {The convergence of bit-wise CID for three-dimensional LJ fluids at various densities. The dashed lines are linear fits to the 3 largest system sizes.}}
\end{figure}

{We also consider the convergence rate of the bit-wise CID $A_{ex}$ as shown in Fig.\ref{fig:finite_size}. The convergence rate is very slow, scaling as $\log_2(\log_2 N)/\log_2 N$, as also observed by Martiniani \textit{et al.}~\cite{Martiniani2019}. Due to the large error bars at smaller system sizes, we only use the 3 largest system sizes to estimate the extrapolated values. However, the uncertainty in the extrapolated values is large.}

\subsection{Cyclically Sheared Binary Lennard-Jones Mixture}

Our next three-dimensional model system is a 80:20 binary mixture of Lennard-Jones particles subjected to athermal cyclic shear. The particles interact via a shifted and truncated Lennard-Jones potential:
% \begin{widetext}
\begin{equation}
U_{\alpha\beta}(r) =
\begin{cases}
& \hspace{-2ex} 4\epsilon_{\alpha\beta}\left[\left( \frac{\sigma_{\alpha\beta}}{r}\right)^{12} - \left(\frac{\sigma_{\alpha\beta}}{r}\right)^6\right] \\
& \hspace{-1ex} \quad  + 4\epsilon_{\alpha\beta}\left[c_{0\alpha\beta} + c_{2\alpha\beta}\left(\frac{r}{\sigma_{\alpha\beta}} \right)^2\right], \; r < r_{c\alpha\beta}\\
& \hspace{-2ex} 0, \hspace{32ex}  r \geq r_{c\alpha\beta}
\end{cases}
\end{equation}
% \end{widetext}
where $\alpha,\beta \in \{A,B\}$ represent the particle type, $r_{c\alpha\beta} = 2.5\sigma_{\alpha\beta}$ is the cut-off distance, and the parameters $\epsilon_{AB}/\epsilon_{AA} = 1.5$, $\epsilon_{BB}/\epsilon_{AA} = 0.5$, $\sigma_{AB}/\sigma_{AA} = 0.8$, $\sigma_{BB}/\sigma_{AA} = 0.88$. The constants $c_{0\alpha\beta}$ and $c_{2\alpha\beta}$ are chosen such that the potential and the forces decay smoothly  to zero at $r = r_{c\alpha\beta}$. The energy and length are in the units of $\epsilon_{AA}$ and $\sigma_{AA}$ respectively.

The initial configurations are obtained from equilibrium liquid configurations by performing an energy minimization. These inherent structures are then subjected to periodic shearing, a cycle of which involves applying the strain sequence: $0$ $\to$ $\gamma_{max}$ $\to$ $-\gamma_{max}$ $\to$ $0$, for a given strain amplitude $\gamma_{max}$. The deformations are performed through an athermal quasi-static (AQS) protocol where an affine transformation of co-ordinates $x' = x + \mathrm{d}\gamma \times z$; $y' = y$; $z' = z$ is applied with $|\mathrm{d}\gamma| \ll 1$, followed by an energy minimization. The shearing is done until the system reaches a steady state. For small strain amplitudes, the stress is proportional to the strain and the system is characterized by localized particle rearrangements or avalanches. Above a critical strain amplitude $\gamma_{y}$, there is a drop in the stress as the system no longer recovers from applied stress, {\it i.e.}, the avalanches are system spanning. This is identified as the yielding point.

\begin{figure*}
\centering
\includegraphics[width=\linewidth]{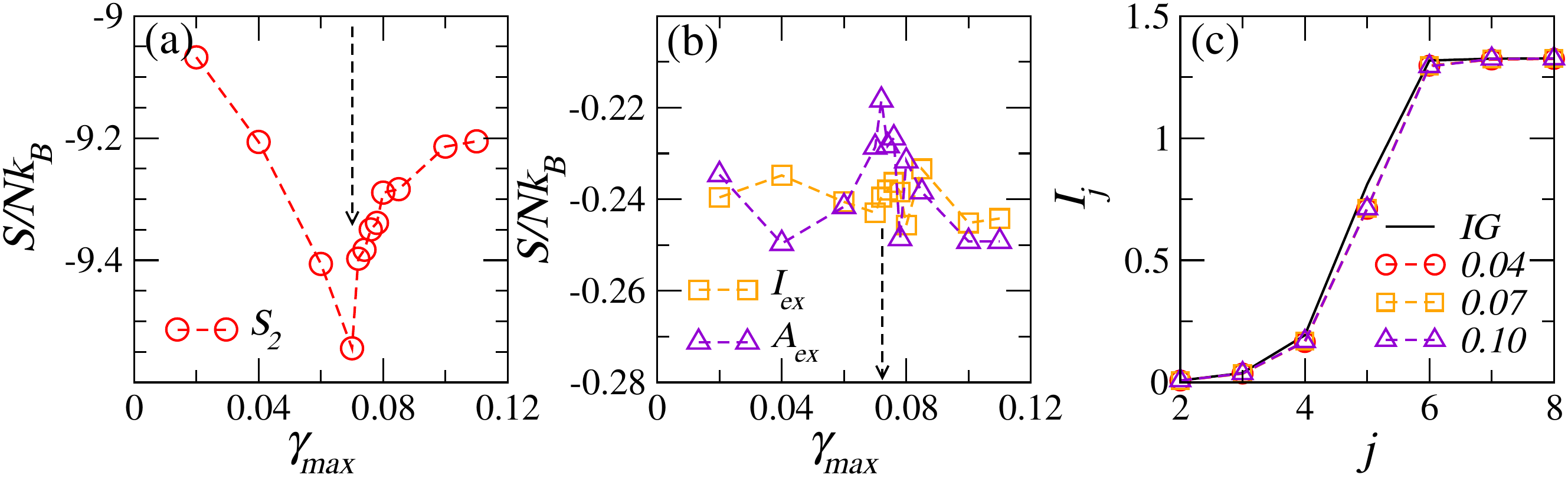}
\caption{(a) and (b) show $S_2$, $I_{ex}$ and $A_{ex}$ for the cyclically sheared binary LJ system as a function of the maximum shear amplitude $\gamma_{max}$. The arrow denotes the yielding point $\gamma_{y} = 0.072$~\cite{Leishangthem2017}. (c) shows the $I_j$ vs. $j$ curves for different $\gamma_{max}$, and IG denotes the ideal gas reference.}
\label{fig:blj}
\end{figure*}

In the following, we present our analysis of the simulation data of Leishangthem et al.\cite{Leishangthem2017}, for the system size $N=64000$ (51200 $A$ and 12800 $B$ particles), number density $\rho = 1.2$, and initial configurations prepared from liquid states at a reduced temperature $T=1$. The yielding point for this system is $\gamma_{y} = 0.072$~\cite{Leishangthem2017}. In Fig.~\ref{fig:blj} (a) and (b) we show the pair correlation entropy $S_2$ and the CID measures for the cyclically sheared binary Lennard-Jones mixture, for various strain amplitudes. Below the yielding point, $S_2$ decreases with $\gamma_{max}$. At the yielding point, we observe a sharp jump in $S_2$. This is in line with our expectation, as the states closer to yielding are more annealed. The CID measures however do not show any clear signal of this transition. This is yet another indication that the different measures of non-equilibrium entropy may yield very different results.  In Fig.~\ref{fig:blj}(c) we show $I_{j}$ vs. $j$ of the particle co-ordinates for different strains $\gamma_{max}$. As seen, the curves for the different strains do not differ much. % (Note by Arun: Why is this so? This is a fairly large system. My initial theory was that the changes in the structure are so small that the discretization procedure used for the grid based CID doesn't capture these differences. But, even the scale-free CID does not show any reasonable signature of the transition.

%A possible reason why the  CID method fails to identify the yielding transition is that the structural changes near the yielding transition are, in fact, very small, as is illustrated by behavior of the pair correlation functions shown in Fig.~\ref{fig:s5}.

\subsection{Cyclically Sheared Binary Soft-Sphere Mixture}
Very similar results are obtained for an equimolar, binary mixture of soft spheres subjected to athermal cyclic shear. The interaction potential is given by the harmonic version of the potential given in Eqn.~\ref{eq:Harmonic}
\begin{equation}
U_{\alpha\beta} =
\begin{cases}
\frac{1}{2}\epsilon\left(1 - \frac{r}{\sigma_{\alpha\beta}}\right)^2, & r < \sigma_{\alpha\beta} \\
0, & r > \sigma_{\alpha\beta}
\end{cases}
\end{equation}
where, $\alpha, \beta \in \{A,B\}$ represent the particle type, $\epsilon = 2.0$ is the strength of the interactions, $\sigma_{\alpha\beta} = (\sigma_{\alpha} + \sigma_{\beta})/2$, $\sigma_{\alpha}$ is the diameter of the $\alpha$ particle, and $\sigma_{B} / \sigma_{A} = 1.4$. The lengths are in units of $\sigma_A$. Depending on the packing fraction $\phi$, the initial states for the shearing are prepared differently. For $\phi$ less than the isotropic jamming density $\phi_j = 0.648$~\cite{Chaudhuri2010}, they are obtained by compressing equilibrium hard sphere configurations at $\phi = 0.363$, through Monte Carlo simulations. For $\phi > \phi_j$, jammed configurations at $\phi_j$ are first obtained using the procedure described in~\cite{Chaudhuri2010}. These are then compressed up to the desired density followed by an energy minimization. The shearing is done through the AQS protocol described in the previous subsection, until a steady state is reached.

For densities below $\phi_j$, the system shows three different states as a function of the strain amplitude $\gamma_{max}$. For small values of $\gamma_{max}$, the system shows point reversibility, {\it i.e.}, the path traced by the system in the configuration space during the backward shear is the same as that of the forward shear. At intermediate strain amplitudes, the system shows loop reversibility, {\it i.e.}, the system comes back to the same configuration at the end of every shear cycle. For larger strain amplitudes, the system goes to an irreversible diffusive state. For $\phi > \phi_J$, the system undergoes yielding at a critical $\gamma_{max}$, similar to the case of the binary LJ mixture. We analyse here two packing densities, $\phi = 0.627$ and $0.72$, corresponding to below and above the jamming density, the configurations for which are obtained from the simulations of Das et al.~\cite{Das2019}. The system size is $N=2000$ particles, and data has been averaged over $9$ independent runs.

\begin{figure*}
\centering
\includegraphics[width=\linewidth]{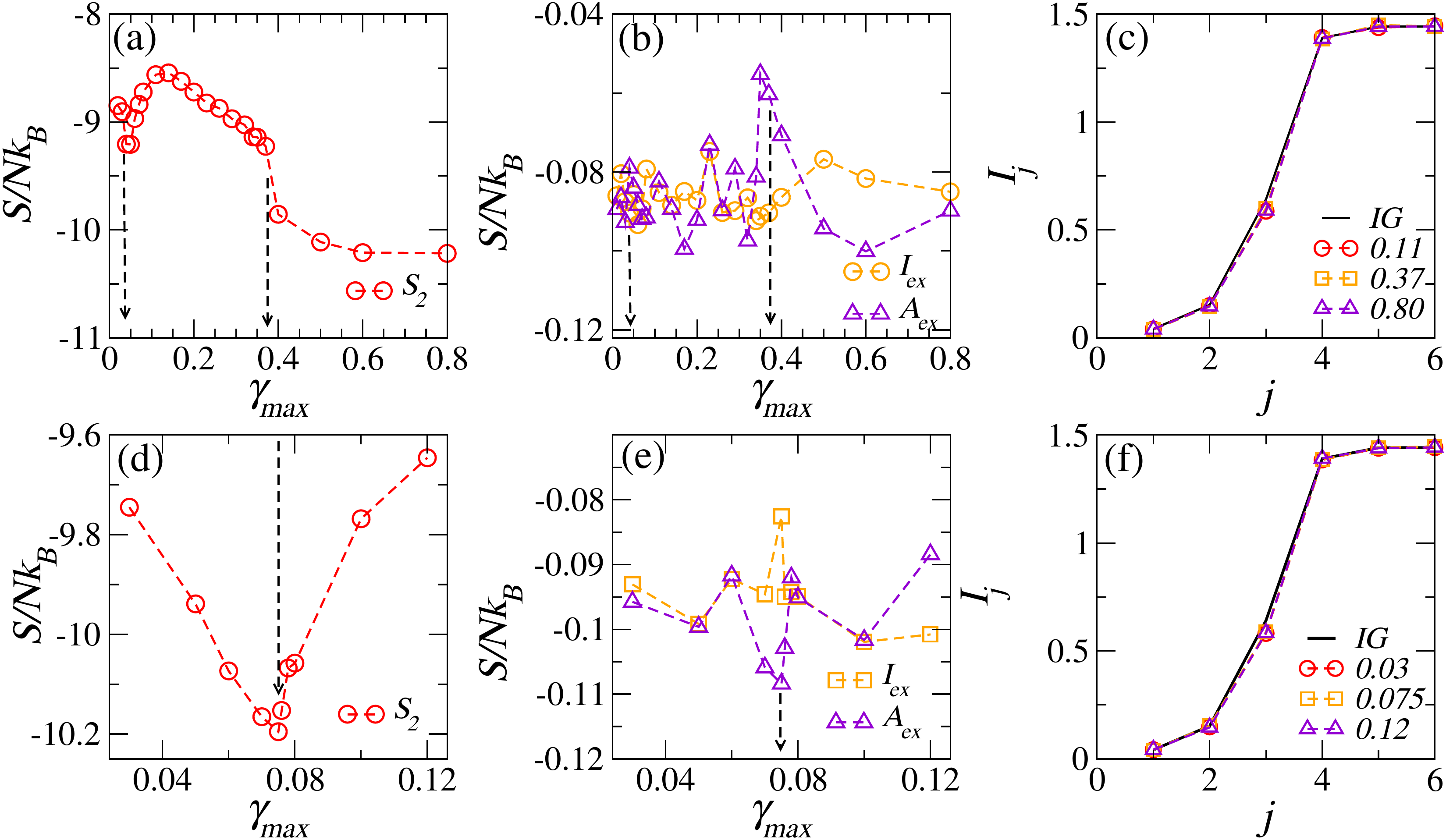}
\caption{$S_2$ and the various CID estimates for cyclically sheared soft spheres as a function of $\gamma_{max}$, for $\phi = 0.627$ (top row) and $\phi = 0.72$ (bottom row). {The arrows denote the various transition points (see text). (c) and (f) show the corresponding $I_j$ vs. $j$ curves for various $\gamma_{max}$, and IG denotes the ideal gas reference.}}
\label{fig:css}
\end{figure*}

Fig.~\ref{fig:css} shows $S_2$ and the various CID estimates evaluated in the steady state as a function of $\gamma_{max}$, for $\phi = 0.627$ (top panel) and $\phi = 0.72$ (bottom panel). It must be noted, however, that the $g(r)$ curves below the jamming density show a power-law divergence at $r=\sigma$ ~\cite{Das2019}.
%(see Fig.~\ref{fig:cssdivergence}). 
This implies that $S_2 = -\infty$ according to Eq.~(\ref{eq:s2}). Nevertheless, we show the results of $S_2$ computed using a fixed bin-width of $0.02\sigma$ for the $g(r)$ calculations. For $\phi = 0.627$, $S_2$ shows a dip around $\gamma_{max} \approx 0.03$ corresponding to the transition from point reversible to loop-reversible state. For $\gamma_{max} > 0.38$, the system transitions into an irreversible state, and a corresponding drop in $S_2$ is seen. The fluctuations in $I$ prevent us from making any statements near the absorbing transition at $\gamma_{max} = 0.03$.
For $\phi = 0.72$ (bottom panel), $S_2$ shows a minimum corresponding to the yielding point at $\gamma_{max} = 0.075$. Again, the CID measures do not show any clear signal of this transition. Similar to the cyclically sheared binary Lennard-Jones system, the bit-wise compression curves for the soft-spheres [Figs.\ref{fig:css}(c) and (f)] do not show much variation for the different strain amplitudes.

\section{Summary and Conclusions}
\label{sec:summary}

We have investigated the use of computational information density as a quantitative measure of the entropy of a variety of equilibrium and out-of-equilibrium systems. The performance of the methods based on data compression varies from reasonable to miserable: In the simplest off-lattice example, \textit{viz.}, the equilibrium one-dimensional hard rod gas, the CID measures perform fairly well in capturing the entropy of the systems. This is perhaps not surprising as the coordinates of a one-dimensional systems map naturally on a linear string, unlike higher-dimensional systems. % For two and three-dimensional systems in equilibrium, the CID measures quantitatively very poor, but  qualitatively ({\em i.e.} after {\em ad hoc} rescaling) in fair agreement with the thermodynamic excess entropy.
However, for non-equilibrium systems - in particular driven systems, there is little agreement between structure-based and information based entropy estimates.

The structure based entropy, seems to capture the ordering transitions in all the cases studied. This prompts further investigation of structural correlation based entropy measures in out of equilibrium conditions. 
%However, as can be seen from the results of the Manna model,  it is questionable if  this measure of (dis)order should be viewed as an out-of-equilibrium entropy.
In the case of driven system, the differences in the order are very subtle and are apparently not captured by either of the information density measures studied here. 

\textmark{It is interesting to compare the predictions of the bit-resolved CID ($A$) with the grid-based version ($I$).
A crucial difference between the computation of $I$ and $A$ is that when we compare $I$ with thermodynamic or structure based entropies is that the grid-based $I$ is only known up to a constant that is not known {\em a priori}, and is therefore determined by fitting. In contrast, there is no such unknown offset when we compare the parameter-free $A$ with the thermodynamic data. Not surprisingly, in the latter case,  the agreement is often less good than that for the (fitted) $I$.} 

\textmark{Whilst it is not surprising that the fitted results agrees better than a procedure with no adjustable parameters, there may be other reasons why the bit-wise procedure is not in agreement with the thermodynamic data.  There might, for instance, be other ways to extract an excess entropy from the bit-wise compressed data.}

\textmark{Also, as we discuss below, whilst the LZ data compression algorithm that we use is lossless, it does build up its ``dictionary'' using a finite sliding window. It is conceivable that the width of this window affects the compression results. However, we have not tested this systematically.} 
%Before closing, we discuss briefly the bit-resolved CID $A$ that we have introduced, and its comparison with the grid-based CID $I$. In the results presented here, a significant difference in the grid-based CID $I$ and the bit-wise CID $A$ is that in the former case, we choose an arbitrary scale factor, in comparing with thermodynamic or structure based entropies, with the rationale that the discretization involved in computing the grid based CID results in an arbitrary scale factor, to compute which we do not have a prescription. On the other hand, the bit-wise CID $A$, should not suffer from such ambiguity, since we do not apply arbitrary discretization. We test to which extent this is so, by comparing $A_{ex}$ values without any scale factor applied. This results in an apparent poorer comparison of  $A_{ex}$ with entropy in some cases, which should be properly interpreted. 
%With the bit-resolved CID, although we expect that there is no loss of information, we cannot rule out the possibility that the procedure for obtaining the aggregate infomation density we have employed (of computing the area $A$) is not optimal. Whether there are better procedures is an important question for future investigation.

\textmark{With these caveats in mind, we can make the following broad statements: (i) For equilibrium systems, both the grid-based and bit-wise CID measures show reasonable agreement with the thermodynamic entropy. % The grid based CID performs better in certain cases (1D Hard Rods and 2D soft-sphere melting at higher densities), whereas the bit-wise CID performs better in others (2D soft-sphere melting at lower densities and 3D Lennard Jones fluid).
(ii) For non-equilibrium systems, which are not driven externally, both the grid-based and bit-wise CID measures show only qualitative agreement with structure based entropy, in capturing the various non-equilibrium transitions. 
% Here, the bit-wise CID performs better for the Manna model, whereas the grid-based CID performs better for 2D Random organization and 2D soft-sphere jamming ([AB:] this is placeholder as we have not seen $S_2$ for jamming)
(iii) For non-equilibrium systems which are externally driven (cyclically sheared binary soft-sphere and binary Lennard-Jones systems in 3D), both the methods perform equally poorly, in the sense that they fail to capture the entropy changes as observed by $S_2$.
(iv) While there are cases where the grid-based CID performs better than the bit-wise CID (and vice-versa), neither method is significantly better than the other, be it that $I$ requires an arbitrary scale factor.
%(v) We do not observe any strong patterns in their performance that could allow us to predict the better method for the different cases such as equilibrium/non-equilibrium, lattice/off-lattice, different spatial dimensions, \textit{etc}.
}

\textmark{We also note a practical limitation of the data compression approach: 
%A few additional considerations must be noted in considering CID as an approach for quantitative estimates of entropy. 
%Apart from the poor quantitative performance of CID in higher-dimensional off-lattice systems, there is also the issue of efficiency of the compression algorithm.
the LZ77 algorithm that we use here is quite slow for long input strings, meaning that for equilibrium systems it would not be faster than other available methods for entropy calculation. 
%(AB: The slowness of the algorithm is probably not a great concern, at least for the kind of system sizes we are looking at. We could remove this statement.)}
While the LZ77 algorithm is supposed to converge to the Kolmogorov complexity in the thermodynamic limit, the rate of this convergence is quite slow ($\sim \log_2 \log_2 N / \log_2 N$) (see, {\it e.g.}~\cite{Martiniani2019}). This implies that the system sizes that we consider here may not be sufficiently large to observe convergence. Moreover, as we mentioned above, any practical implementation of the LZ77 algorithm (and most other compression methods) uses a sliding window -- a fixed-length buffer that slides over the given symbolic string -- within which it computes the LZ77 factorization. This is necessary to keep the memory requirements and the computational cost reasonable. Consequently, such an implementation while lossless, would not give the correct value of the complexity for very long strings. 
The effect of the finite window size may also limit the use of CID as a quantitative entropy estimator. 
%These considerations must also be taken into account when interpreting the CID results. 
}

%\DFcomment{What is the message here? Are you saying: 1. In principle, compression should work, but in practice LZ does not keep enough information in memory or 2. We are too far from the themrodynamic limit?} {\color{magenta}([AB:] It is a combination of both these considerations. I should have mentioned that this could be one of the reasons why $A_{ex}$ does not perform so well as the $A_{id}$ I have rewritten the paragraph above.)}\\

%{\bf \color{red} SS: Arun, mention of the 'scaling arguemnt' is unclear above. Also, the referee asked for a comparison of I and A. Should state it a bit better. I guess the data not shown is not mentioned anymore? Reg. DF remark above}

\textmark{These considerations suggest that the CID measures studied here may, 
at this stage, not offer competitive approaches to estimate non-equilibrium entropies. However, the methods do offer an interesting perspective on the relation
%not be appealing as practical methods to compute entropy. Rather, their investigation is attractive as an approach to exploring the interesting relationship 
between information and entropy in physical systems.} Thus we conclude that, while enticing, the use of information based entropy measures -- particularly attractive in nonequilibrium contexts -- requires further insights and developments to be employed as a useful tool in understanding the nature of order and entropy in equilibrium and nonequilibrium situations. 
 
% {\color{magenta} (Note by Arun) I think I understand why the LZ77 algorithm performs objectively worse at very small bin-widths, where the length of the string is very large. For practical reasons, the LZ algorithm cannot keep track of all the unique words in the entire string, because that would require checking if the current buffer of symbols matches with any of the previously encountered words. For this reason, the LZ77 algorithm implements a ``sliding window'', i.e., it will only check for matches within a certain length preceding the buffer's current position. It should be clear enough to see why, for very large strings, the method will not achieve perfect compression: some words might need to be added to the dictionary again even though they have been encountered previously, because the window has slid beyond its previous occurrence. This seems to introduce quite a significant amount of noise in the complexity or the number of LZ77 factors. I have not figured out what is the length of the sliding window of the LZ77 algorithm that we are using. I will need to do some more digging. But what could be the solution to this problem? It could be cute if we can raise the size of this sliding window, but the resulting increase in the computational cost might be a problem. Or we can try some other method which does not involve a sliding window, and which is known to converge to the Shannon-limit. Surprisingly, there seems to be no mention of this limitation in the articles that have used the LZ77 for their analyses. Or have I missed this?}

\section*{Acknowledgments}

The authors would like to thank in particular Stefano Martiniani for much practical help, and also  Dov Levine, and Roy Beck-Barkai for useful discussions. The authors acknowledge support from UKIERI-DST under Grant No.s IND/CONT/G/16-17/104, DST/INT/UK/P-149/2016. This research was supported in part by the National Science Foundation under Grant No. NSF PHY-1748958, and the hospitality at KITP, UCSB, under the program ``Memory Formation in Matter'' is gratefully acknowledged by DF and SS. We also thank the International Centre for Theoretical Sciences (ICTS) for hospitality and support during the program ``Entropy, Information and Order in Soft Matter'' ICTS/eiosm2018/08. SS acknowledges support through the J C Bose Fellowship (DST, India). 

\section*{References}

\bibliography{entropy-id}

\end{document}